\documentstyle[aps,prl,epsf,amssymb,multicol,float]{revtex}
\begin{document}
\title{Anomalous fluctuations of active polar filaments} 
\author{Tanniemola B. Liverpool}
\address{
  Condensed Matter Theory Group, Blackett Laboratory, Imperial
  College, London SW7 2BZ, U.K. \\
Kavli Institute for Theoretical Physics, University of California, Santa Barbara, CA 93106 } \date{\today}
\maketitle
\begin{abstract}
  Using a simple model, we study the fluctuating dynamics of
  inextensible, semiflexible polar filaments interacting with active
  and directed force generating centres such as molecular motors.
  Taking into account the fact that the activity occurs on time-scales
  comparable to the filament relaxation time, we obtain some
  unexpected differences between both the steady-state and dynamical
  behaviour of active as compared to passive filaments. For the
  statics, the filaments have a {novel} length-scale dependent
  rigidity.  Dynamically, we find strongly enhanced anomalous
  diffusion.
\end{abstract}


\begin{multicols}{2}
    Filamentous proteins are major components of the cell
  cytoskeleton\cite{howard}.  Examples are actin filaments,
  microtubules and intermediate filaments. Their mechanical properties
  are important for cell stability and support and have been well
  studied at thermodynamic equilibrium by {\em in vitro} experiments.
  However, the conditions in the living cell are very different from
  those in the laboratory. Protein filaments interact with other
  proteins such as molecular motors and cross-linkers.  This has led
  to a flurry of recent {\em in vitro} experiments of mixtures of
  filaments and their associated proteins in order to compare with
  their purified state~\cite{nedelec,kas,loic}. From a theoretical
  point of view, the proteins are typically far from equilibrium and
  therefore even to understand their steady-state behaviour, one has
  to study their dynamics.  Non equilibrium effects have also recently
  been studied in biological membranes~\cite{act_mem}.
  
  In this paper, motivated by recent experiments on F-actin and
  myosin in the presence of ATP~\cite{loic}, we study one example of
  the non-equilibrium behaviour of bio-filaments: the fluctuating
  dynamics of polar filaments with active centres.  This is also a
  model system for the study of non-trivial aspects of semiflexible
  filament dynamics.  A key point of our analysis is the fact the
  activity of the proteins occurs over a time-scale $\tau$ which may
  be comparable to the relaxation time of the filament. Unlike recent
  work on motile solutions~\cite{motile}, the active centres
  considered here are associated with single filaments (i.e.  are not
  cross-links) and cannot move one filament with respect to another.
  The viscosity of the solvent is given by $\eta$.  We note that the
  typical energy scale of a biochemical reaction is of the order of a
  few $k_B T$ at physiological conditions.
 
\begin{figure}
  
\centerline{\epsfxsize=8truecm \epsffile{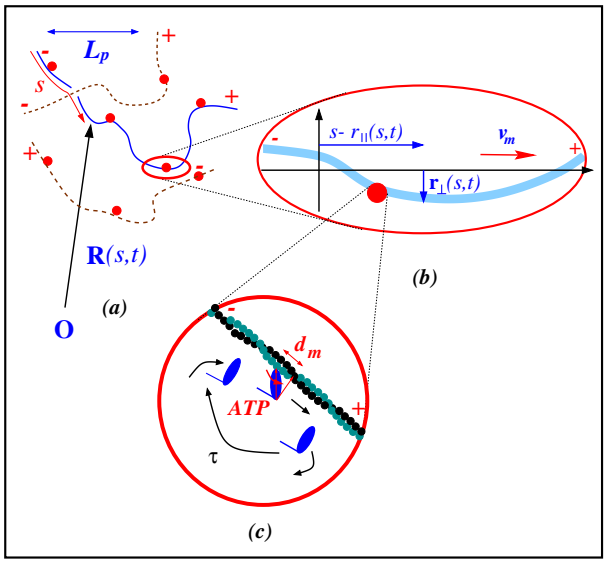} }
  \vskip 0.5cm
\caption{ \protect (a) The fluctuating {\em polar} filaments of 
  persistence length $L_p$ parametrised by ${\bf R}(s,t)$ decorated by
  active centres. Note that there is a $(+)$ and a $(-)$ end for each
  filament. (b) The filament on length-scales below $L_p$ showing the
  transverse ${\bf r}_{\perp}(s,t)$ and longitudinal $r_{\|}(s,t)$
  motion. (c) A schematic of the cycle of activity of a motor such as
  myosin with activity time $\tau$.  }
\label{fig:filament}
\end{figure}

  The filament can be parameterised by a curve through its centre,
  ${\bf R}(s)$ (see Fig.~\ref{fig:filament}). The unit tangent-vector
  is defined as ${\bf t}(s)= \partial {\bf R} / \partial s$. An easily
  measured quantity using e.g., video microscopy is the steady-state
  tangent correlation function $C_{\bf t}(s) \equiv \langle {\bf t}(s)
  \cdot {\bf t}(0) \rangle$. For a semiflexible polymer in equilibrium
  it decays exponentially, $C_{\bf t}(s) = \exp \left[- |s|/L_p
  \right]$ which defines the persistence length $L_p$.

\paragraph*{Main results} We calculate $C_{\bf t}(s)$ and find that 
due to activity on a time-scale $\tau$, the filaments develop a {\em
  length-scale} dependent bending rigidity.  On short length-scales,
typical conformations have the bare persistence length whilst on
longer length-scales the filaments may be characterised by a lower
`renormalized' persistence length.  There is a {\em cross-over} length
between the two regimes, $\ell_c \sim \left( \tau k_B T L_p / 2 \pi
  \eta \right)^{1/4}$.  Therefore an analysis of the filament
conformations can give information about the time-scale $\tau$ at
which the activity occurs.  We also obtain modified relaxational
dynamics with anomalous diffusion which we show has an effect on the
high frequency viscoelasticity. 
The shear modulus of the active filament solution apparently has a
frequency dependent effective temperature.
There is a crossover from at high frequencies a shear modulus
corresponding to the bare temperature to at lower frequencies a
modulus equivalent to the higher renormalized temperature both with a
scaling of $\omega^{3/4}$. In the cross-over regime, the modulus {\em
appears} to have a stronger frequency dependence with a power law
$G^*(\omega) \propto \omega^{\alpha}$ where $\alpha > 3/4$. A simple
consequence of filament polarity is a {\em small} ballistic 
component to the motion of the filament,
  
\paragraph*{Explicit analysis} 
We model the filaments with the Kratky-Porod worm-like
chain~\cite{doieds,isamag} which takes into account the {\em bending}
energy cost of the chain. The worm-like chain Hamiltonian is
\begin{math}
{\cal H}_{wlc}[\{{\bf R}(s)\}] = 
{\kappa \over 2 }\int_{-L/2}^{L/2} ds \left({\partial^2 {\bf R} \over \partial s^2} \right)^2 . 
\end{math}
The stretching energy of chain molecules is much higher than the
bending energy and we may consider the chain as inextensible.  The
persistence length is defined as
\begin{math}
  L_p=\kappa / k_B T
\end{math}.

The dynamics at finite $T$ may be expressed by the Langevin equation
\end{multicols}
\begin{eqnarray}
  {\partial\over \partial t} {\bf R} (s,t) &-&
  \int ds' \, \, {\bf H}(s,s') 
  \cdot
  \left[- \kappa {\partial^4 {\bf R} \over \partial s'^4} +  {\partial \over \partial s'} \left( \Lambda (s',t) 
{\partial {\bf R} \over \partial s'} \right) \right]- {\bf V}(s,t) ={\bf f}
(s,t)  + {\bf f}^{(m)}
(s,t)  , \label{langevin} \end{eqnarray} 
\begin{multicols}{2}
  where $\Lambda(s,t)$ (a Lagrange multiplier) is an {\it
    instantaneous}, fluctuating `tension' which enforces the {\em
    local} inextensibility of the chain.  The inextensibility
  constraint fixes the tension, $\Lambda(s,t)$, by 
\begin{math}
  0 = \left|{\partial {\bf R} \over \partial s}\right|^2 - 1 \, .
  \label{constraint}
\end{math}
Because of this constraint, the dynamics of semiflexible filaments is
non-linear and generally insoluble.
In addition to the thermal velocities ${\bf f} (s,t)$, there is an
additional active velocities ${\bf f}^{(m)} (s,t)$ correlated over a time
$\tau$.  The polarity of the filament implies the active forces have a
mean direction along the contour of the filaments (see Fig
\ref{fig:filament}).  Hydrodynamics is taken into account by the
mobility tensor ${\bf H}(s,s')$. In the Stokes approximation valid for
low Reynolds number, ${\bf H}(s,s')= {\bf H}({\bf R}(s)-{\bf R}(s'))$
and generates a non-dissipative term~\cite{Chaikin}, ${V}_i(s,t)= k_B
T \int_0^L d s'\delta H_{ij} (s,s')/ \delta R_j(s',t) $. In addition
${\bf f}(s,t)$ will also depend on ${\bf R}(s,t) $~\cite{Chaikin}. We
assume that the activity does not affect the thermal forces, i.e. the
active proteins do not affect the collisions of the solvent molecules
with the filament. Then the thermal velocities satisfy the
fluctuation-dissipation theorem, and
\begin{math} {\bf f}(s,t)
\end{math}  have zero mean and Gaussian fluctuations,
given~\cite{Chaikin} by \begin{math}\langle f_i(s,t)
  f_j(s',t') \rangle = 2 k_B T H_{ij} (s,s') \delta(t-t') 
\end{math}. 

We consider rod-like segments and restrict the analysis to
length-scales $\ell$ below $L_p$ so that we can decompose the dynamics
into transverse and longitudinal motion (see Fig. \ref{fig:filament})
and write the position of the filament as ${\bf R}(s,t)=
(s-r_{\|}(s,t))\hat{\bf u}(t) + {\bf r}_{\perp}(s,t)$ where $s \in \{-
\ell/2,\ell/2 \}$ and $\hat{\bf u}$ is a time-dependent unit vector
giving the orientation.  We can obtain results within a systematic
small gradient expansion for $|{\bf r}_{\perp}(s,t)|, |r_{\|}(s,t)|
\ll s$.  The mobility tensor is given by
\begin{math}
  {H}_{ij}(s,s') = h(s-s') \left(\left({\delta}_{ij}-\hat{u}_i\hat {u}_j \right) + 2 \hat{u}_i \hat{u}_j
    + O(|\partial_s {\bf r}_{\perp}|^2) \right)
\end{math} where in this and the following, for a function $A(x)$,
\begin{math}\partial_x A \equiv \partial A/ \partial x
\end{math}. 
We can thus decompose eqn. (\ref{langevin}) into parallel and
perpendicular components in an expansion to $O(|\partial_s {\bf r}_\perp|^2)$,
\end{multicols}
\begin{eqnarray}    
{\partial_t {\bf r}_\perp } (s,t) &=&  \int ds' h(s-s')
\left[ - \kappa  \partial^4_{s'} {\bf r}_\perp +  \Lambda(s',t)    
\partial^2_{s'} {\bf r}_\perp + \partial_{s'} \Lambda \partial_{s'}
 {\bf r}_\perp \right] + {\bf f}_\perp(s,t)  + {\bf f}^{(m)}_{\perp} (s,t) 
 \label{perpdyn} \\
 \partial_t r_{\|} (s,t) &=& \int ds' 2 h(s-s') 
 \left[  - \kappa  \partial^4_{s'} r_{\|} - \partial_{s'} \Lambda   \right] + f_{\|}(s,t) + f_{\|}^{(m)}(s,t)  \label{parrdyn}\, ,
\end{eqnarray}  
\begin{multicols}{2}
  which are coupled by the constraint of inextensibility,
\begin{equation}
\partial_s r_{\|}= {1 \over 2} (\partial_s {\bf r}_\perp)^2 + O(|\partial_s {\bf r}_\perp|^4) \,.\label{inextensible}
\end{equation}
For simplicity in most of this paper, we consider the Rouse model
  which assumes local friction. We also focus on dilute solutions.
  Long range hydrodynamics and non-zero concentration of filaments
  will modify some of the results (see below).  For the Rouse model
  $h(s-s')= \delta(s-s')/ \zeta_{\perp}$ where
  \begin{math} {\zeta_{\perp}} = { 4 \pi \eta } = {2} {\zeta_{\|}}
\end{math}.   We
model the active velocities by a Gaussian noise with mean $\langle {\bf
  f}_{\perp}^{(m)} \rangle = 0 $ and $
\langle f_{\|}^{(m)} \rangle =
v_m $, a drift, reflecting the polar nature of the filament
and mean square fluctuations, $\delta {\bf f}^{(m)} = {\bf f}^{(m)} -
\langle {\bf f}^{(m)} \rangle$ given by
\begin{equation}\langle \delta f_i^{(m)}(s,t) \delta f_j^{(m)}(s',t')
  \rangle = \sqrt{2} {\alpha_i \over \zeta_i^2} \Theta\delta_{ij} \delta(s-s') \Phi(t-t')
\label{act_force}\end{equation} where  
$ \Phi(t) = \exp\{-|t|/\tau\}$ and $\{i,j\}$ refer to $\perp,\|$. The
level of {\em activity} is controlled by the parameter $\Theta$.
Correlations of the active force decay over a time, $\tau$ the typical
activity period.  The constants $\alpha_{i,j}$ measure the relative
partitioning of the activity between transverse and longitudinal
components and satisfy the relationship $\alpha_{||}^2
+\alpha_\perp^2=1$.  In general, the force applied by the motor
protein on the filament will not be purely tangential depending for
example on its trajectory of approach of the filament and the
conformation of the neck and chain region~\cite{howard}. Averaging
over orientations, {\em only} the longitudinal component will have a
non-zero average, because of the filament polarity. In addition, the
remainder of the energy of the actomyosin reaction that is not
converted to directed work will be dissipated as heat contributing to
the fluctuations on the same (reaction) time-scale. It is difficult
apriori to estimate $\alpha_{i,j}$ and we choose somewhat arbitrarily,
$\alpha_\perp=\alpha_{||}=1/
\sqrt{2}$. Between active events the `motors' diffuse freely in the
solution so it is reasonable to assume that there is no spatial
correlation between active sites. We emphasise that unlike the thermal
noise, the active noise correlations {\em do not} satisfy the
fluctuation dissipation theorem.

The eqns.(\ref{perpdyn},\ref{parrdyn},\ref{inextensible}) are most
easily studied by analysing the motion of bending modes of wave-vector
$q$ and frequency $\omega$.  Defining $F(s,t) = \int {dq \over 2 \pi}
{d \omega \over 2 \pi}\tilde{F}(q,\omega) \exp (i \omega t + q s)$ we
only consider $q$ such that $\pi/a \gg q \gg \pi/ \ell$.  On
length-scales below $L_p$, $C_{\bf t}(s)$ can be obtained 
from the transverse fluctuations.  The transverse dynamics are
approximately given by
\begin{equation}
\tilde{\bf r}_\perp(q,\omega)  \simeq  { \tilde{\bf f}_{\perp}(q,\omega)  +  
\tilde{\bf f}^{(m)}_{\perp}(q,\omega)\over i \omega + \alpha q^4 }\, ,
\label{perp-q}\end{equation}
where $\alpha = \kappa/ \zeta_{\perp}$. Corrections due to the tension
$\Lambda(s,t)$ are higher order in the gradient expansion. The
transverse fluctuations at time $t$ are relaxed over a length-scale
$\ell_\perp(t) = (\alpha t)^{1/4}$. Because of inextensibility, eqn.
(\ref{inextensible}) there is an induced {\em time-independent}
longitudinal motion $r_{\|}^{(0)}(s)$ due to the averaged transverse
motion, $\partial_s r_{\|}^{(0)}(s) = \left\langle {1 \over 2}
  (\partial_s {\bf r}_\perp)^2 \right\rangle_{\perp} $.

\begin{figure}
  
\centerline{\epsfxsize=8truecm \epsffile{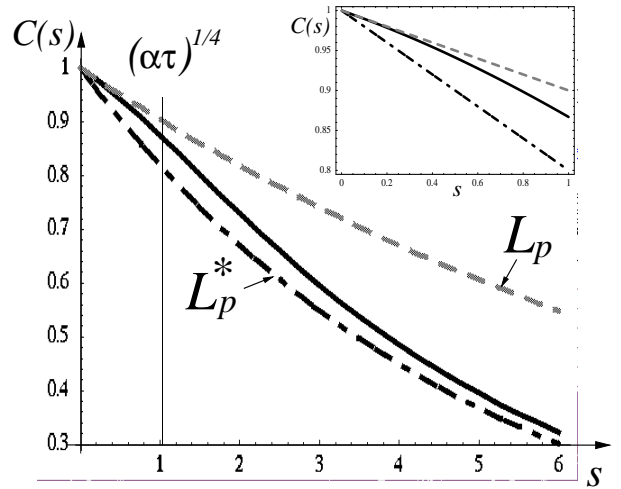} }
  \vskip 0.5cm
\caption{ \protect The tangent correlation function for the active filament of {\em bare} persistence length $L_p=10$, compared with those simple wormlike chains of persistence length of $L_p=10$ and $L^*_p=5$ using units such that $\alpha \tau =1$. Inset: $C_{\bf t}(s)$ near $s=0$. }
\label{fig:tancor}
\end{figure}

The tangent correlation function is given by 
\end{multicols}
\begin{equation} C_{\bf t}(s) = 
\left \langle {\bf t}(s,t)  \cdot {\bf t}(0,t) \right\rangle \simeq  \left \langle 1-  \partial_s r_{\|}^{(0)}(s,t)  - \partial_s r_{\|}^{(0)}(0,t)  + \partial_s {\bf r}_{\perp} (s,t) \cdot \partial_s {\bf r}_{\perp} (0,t)  \right\rangle_{\perp} \, .
\label{tan_cor}\end{equation}
\begin{multicols}{2}
  Using eqns. (\ref{inextensible},\ref{perp-q},\ref{tan_cor}), the
  tangent correlation function has the form
\begin{equation}C_{\bf t}(s) \simeq \exp \left\{ -
{s \over {L_p}} - {2 \Theta\over \alpha^2} \int {dq \over 2 \pi} {1- \cos (qs) \over q^2 ( q^4 + (\alpha \tau )^{-1})} \right\} .
\end{equation}
The correlation function is plotted in Fig. \ref{fig:tancor}.  For
length-scales less than $\ell_c \simeq \ell_\perp(\tau) = \left(\kappa
  \tau / \zeta_\perp^{} \right)^{1/4}$ the effective persistence
length is approximately equal to the bare persistence length $L_p$
whilst for length-scales above $\ell_c$, the conformations can be well
modelled by a worm-like chain with a lower renormalized persistence
length given by $L^*_p= L_p \left( 1 + \Theta\tau \zeta_\perp / k_B T
\right)^{-1}$.  Dimensional analysis suggests a new {\em active}
temperature scale given by $k_B T_{a}=\Theta\tau \zeta_\perp$.

\paragraph*{Dynamics} The dynamics of inextensible filaments is
anisotropic~\cite{everaers,morse,livmag,morse01}: longitudinal motion
has different relaxation dynamics to the transverse.  Viscous
dissipation due to longitudinal motion must be taken into account,
giving a time-scale over which `tension' propagates along the
filament~\cite{everaers,morse,livmag,morse01}. From eqn.
(\ref{perp-q}), we can calculate the transverse dynamic fluctuations,
$R_{\perp}^2(t)=\left\langle ({\bf r}_{\perp}(s,t) - {\bf
    r}_{\perp}(s,0))^2 \right\rangle_{\perp}$
\begin{equation}
R_{\perp}^2(t) = \left\{\begin{array}{lr} t^2 {\Theta \over \sqrt{2}} \kappa^{3/4} \zeta_\perp^{-3/4} \tau^{7/4}, & t \ll \tau \\ 
t^{3/4}{8 \Gamma (1/4) \over 3 \pi
    } \kappa^{1/4}\zeta_\perp^{-3/4} (k_B T + \Theta \zeta_\perp \tau), & t \gg \tau\, \end{array}
\right.
\end{equation}
There is a cross-over from ballistic motion at short times to
sub-diffusive behaviour at long times. Note the thermal and active
contributions to the sub-diffusive regime.

We can solve for the longitudinal motion self-consistently as follows:
we average over {\em only} the transverse fluctuations and use the
inextensibility constraint~\cite{livmag,morse01} to give a
relationship between the tension and the longitudinal motion, $i q
(\tilde{r}_{\|}(q,\omega) - \tilde{r}_{\|}^{(0)}(q)) = K(\omega)
\tilde{\Lambda}(q,\omega)$ defining a frequency dependent {\em
  extensional} compliance, $K(\omega) = K_{eq}(\omega) + K_{a}(\omega)$
which has equilibrium and active contributions. The equilibrium
modulus is given by $K_{eq}(\omega) \simeq 2^{-3/4} k_B T
\kappa^{-5/4} \left(i \zeta_\perp \omega \right)^{-3/4} $ is not new
and has been obtained previously~\cite{morse,freds}. We find in
addition a {\em new} active contribution given by
\begin{equation} K_{a}(\omega) = 
{4 \Theta\tau \over \kappa} \int_k { (\alpha k^4 \tau + f(\alpha k^4 \tau) - i \omega \tau )\over f(\alpha k^4 \tau) (2 \alpha k^4 - i \omega) (f(\alpha k^4 \tau)- i \omega \tau) } \label{k_act}
\end{equation} where $f(x)=x +1$.
In the limit $\omega \rightarrow 0$, it is given by $K_{a}(\omega)
\simeq 2^{-3/4}{\Theta\tau \zeta_{\perp} \kappa^{-5/4} }\left[\left( i
    \omega \zeta_\perp \right)^{-3/4} - {1 \over 4} \left(\zeta_\perp /
    2\tau\right)^{-3/4}\right] + O(\omega)$. The modulus $K(\omega)$ is
then substituted into the longitudinal dynamics.  This self-consistent
approach corresponds to an infinite resummation of a set of diagrams
of the perturbation expansion in $\Lambda$\cite{livmag}.  We obtain
the following equation for the longitudinal motion,
\begin{eqnarray}
\tilde{r}_{\|}(q,\omega)  - \tilde{r}_{\|}^{(0)}(q)& \simeq & 
{ \tilde{f}_{\|}(q,\omega) + \tilde{f}^{(m)}_{\|}(q,\omega) 
\over i \omega + q^2 K^{-1}(\omega)/\zeta_{\|}} 
\end{eqnarray} 
As defined above, $ r_{\|}^{(0)}(s)$ is the time-independent motion in
the longitudinal direction due to the averaged transverse motion.
Given this, we calculate the longitudinal dynamical fluctuations,
$R_{\|}^2(t)=\left\langle (r_{\|}(s,t) -  r_{\|}(s,0))^2 \right\rangle_{\|}$
\begin{equation}R_{\|}^2(t)
 = \left\{\begin{array}{lr}  t^2 \left( v_m^2 + \Theta \zeta_{\|}^{1/2} 2^{-3/4} \kappa^{3/8} \zeta_\perp^{-3/8} \tau^{1/8}\right), & t \ll \tau \\ 
t^2 v_m^2 + t^{7/8} {8 \Gamma (1/8)  \over 7 \pi } { (2 
\kappa)^{5/8} (k_B T+\Theta \zeta_\perp \tau)^{3/8}\over \zeta_{\|}^{7/8}}
   & t \gg \tau\,.  
\end{array}\right.
\end{equation}
Because of the drift term (the directed motor/filament interaction),
the longitudinal diffusion is ballistic for all time-scales.

\begin{figure}
  
\centerline{\epsfxsize=8truecm \epsffile{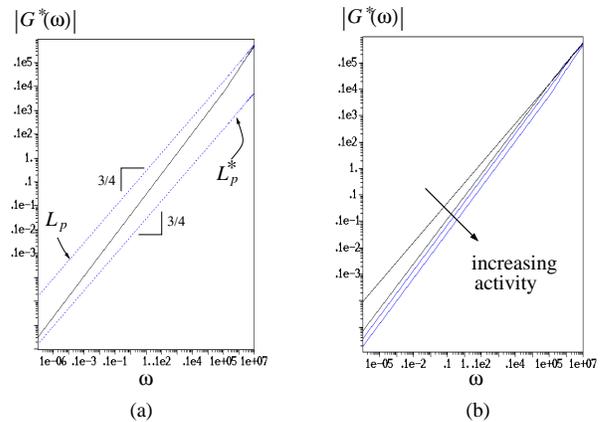} }
  \vskip 0.5cm
\caption{ \protect The magnitude of the high frequency modulus for 
  active filaments.  We have chosen units of length and time such that
  $\alpha \tau = 1$ and $k_B T/ \zeta_{\perp}=1$. In these units, the bare
  persistence length of the filaments is $L_p=1$. (a) The cross-over
  to a renormalised persistence length $L_p^*=1/101$. (b) Increasing
  activity corresponding to $\Theta \zeta_\perp \tau = k_B T_a =
  25,50,100$}
\label{fig:gomega}
\end{figure}

\paragraph*{High frequency viscoelasticity} 
The complex shear modulus of a solution of semiflexible filaments at
high frequencies is dominated by the extensional compliance and
$G^*(\omega)= {2 \over 15} \rho K^{-1}(\omega) \simeq 0.133 {\rho /
  \left(K_{eq}(\omega)+K_a(\omega)\right)}^{}$~\cite{motile,morse,freds}
which for a solution of passive filaments is given by $G^*(\omega)
\propto \omega^{3/4}$. For the active filaments, using eqn.
(\ref{k_act}), we plot the absolute value of the complex high
frequency modulus for varying activity in Fig. \ref{fig:gomega}.  We
see a crossover from at high frequencies a modulus corresponding to
the bare temperature, $T$ to at low frequencies a modulus equivalent
to the higher renormalized active temperature $T+T_{a}$, {\em both}
with a scaling of $\omega^{3/4}$.  In the cross-over regime, the
modulus appears to have a stronger frequency dependence with a power
law $G^*(\omega) \propto \omega^{\alpha}$ where $\alpha > 3/4$. It is
interesting to note that the cross-over occurs over a very wide
frequency range.

\paragraph*{Hydrodynamics}
Within the (screened) Oseen approximation the mobility tensor is given
by~\cite{doieds},
\begin{math} 
{H}_{ij}[{\bf r}] = { e^{- |{\bf r}|/\xi} \over 8 \pi 
  \eta |{\bf r}|} \left({\delta}_{ij}+\hat{r}_i\hat {r_j}
\right) \, 
\end{math} for  $|{\bf r}| > a$.
The hydrodynamic screening length $\xi \rightarrow \infty$ for dilute
solutions and is equal to the mesh-size $\xi = (\rho_a a)^{-1/2}$ for
semi-dilute solutions with actin concentration $\rho_a$ and
filament diameter $a$..  Therefore, $h(s-s') = { e^{- |s-s'|/ \xi}
\over 8 \pi \eta |s-s'|} $ in eqns (\ref{perpdyn},\ref{parrdyn}).  For
dilute solutions ($\xi \rightarrow \infty$), this gives logarithmic
corrections to the Rouse model described above and implies a
modification of the cross-over length to $\ell_c \sim ( \kappa \tau /
\zeta_\perp \log[( \kappa \tau / \zeta_\perp)^{1/4}/a])^{1/4}$. For
concentrated solutions ($\xi$ finite), the friction coefficients
cross-over length is $\ell_c \sim (\kappa \tau / \zeta_\perp\log[\xi/
a])^{1/4}$.
\paragraph*{Entangled solutions}
For entangled solutions, the filament can be modelled as confined in
its {\em tube}. Given a mesh-size $\xi$, we can define a tube diameter
$D_e\sim L_p (\xi/L_p)^{6/5}$ and entanglement length $L_e \sim L_p
(\xi/L_p)^{4/5}$~\cite{semiflex}. The tube can be modelled as a
confining potential for ${\bf r}_{\perp}$ which we can model as
$V_{tube}= {1 \over 2} k |{\bf r}_\perp|^2$ with $k \simeq \kappa /
L_e^4$ chosen so as to give $\langle |{\bf r}_\perp(L_e) - {\bf
  r}_\perp(0)|^2 \rangle^{1/2} = D_e$, the tube diameter.  We obtain
the tangent correlation function,
\begin{math}C_{\bf t}(s) 
= 1- {s \over {L_p}} - {2 \Theta \over \alpha^2} \int {dq \over 2 \pi} {1- \cos (qs) \over q^2 ( q^4 + (\alpha \tau )^{-1} + k/\kappa )}  .
\end{math}

\paragraph*{Discussion} Our results should be relevant for recent
experiments on mixtures of F-actin and S1 Myosin~\cite{loic}.  The
molecular motor myosin interacts with F-actin in the presence of ATP
undergoing a conformational change in the process. Hydrophobic
interactions between tails of a common variant, Myosin II, lead to
multi-headed clusters which can act as active
cross-links~\cite{kas,motile} between two or more filaments.  In
contrast, S1 myosin is a single-headed version of myosin that is
without a tail. Therefore, in general S1 myosin/ATP interacts with
single {\em polar} actin filaments applying biased non-equilibrium
forces. The S1 experiments found surprisingly different
steady-state and dynamic behaviour of the filaments as compared to the
pure F-actin system~\cite{loic}.

We now estimate the values for the parameters of our model
corresponding to the S1 experiment.
For an actin monomer concentration $\rho_a$, the fraction of bound
myosin $\phi$ can be estimated using the equilibrium
constant~\cite{howard}, $k_{eq} = 500 \, \mbox{nM}$ for the passive
reaction ($\mbox{Actin} + \mbox{Myosin}\Leftrightarrow
\mbox{Actomyosin}$). The fraction of bound myosin is given by $\phi =
\rho_a /(k_{eq} + \rho_a)$ leading to the typical separation of the active sites (motors) on the
filament, $\ell_m \simeq (\phi \rho_m
\xi^2)^{-1} = a (\rho_a / \phi \rho_m)$ where $\rho_m$ is the concentration of S1 myosin.  Then we may divide
the filament into regions of size $\ell_m$ in which the motors are
expected to act independently. Therefore $\langle {\bf f}(s,t) {\bf
f}(s',t')\rangle = \ell_m \delta(s-s')\langle {\bf f}_m(t) {\bf
f}_m(t')\rangle$.  We assume that motor attachment on the filaments is
a Poisson process~\cite{gardiner}, i.e. that motors arrive at random
times $t_n$ with a constant rate $\lambda$. The number of motors
arriving in a period $\Delta t \gg 1/\lambda$ has a Poisson
distribution.  The forces applied by each motor are assumed to decay
over a timescale $\tau$ so that the force at time $t$ is given by
\begin{equation}
{\bf F}_m(t) = \sum_{n=1}{ a \over \ell_m}  {\bf f}_0 g(t-t_n)
\end{equation}
where $g(t)= \exp (- t/\tau), t>0$ and $g(t)=0, t <0$ and ${\bf f}_0$
is the typical force applied by a motor to a filament of diameter $a$.
After some standard manipulations~\cite{gardiner} and averaging over
orientations, the velocity correlations can be shown to be given by
eqn. (\ref{act_force}) above with the activity parameter given by
$\Theta \simeq \lambda \tau
\ell_m^{-1} (f_0 a)^{2}$ and  we set 
$\lambda = 1/ \tau$. Similarly, the drift can be estimated as $v_m
\simeq (f_0/\zeta_\|) \ell_m^{-1}$.  The typical force, $f_0$ and
active time-scale, $\tau$ can be obtained from biochemical data.  The
viscosity of water is $\eta
\simeq 10^{-3} \, \mbox{Pa s}$ and F-actin has diameter $a \simeq 7\,
\mbox{nm}$ and persistence length $L_p
\simeq 15 \mu \mbox{m}$. 
The stall force of Myosin is approximately $5 \, \mbox{pN}$ and the
duration of the motor cycle is approximately $\tau \simeq 5 \,
\mbox{ms}$ The motor step-size $d_m$ is approximately $d_m \simeq 10
\,\mbox{nm}$. 
For ATP saturation and $5 \mu \mbox{M}$ S1 and $14\mu \mbox{M}$
F-actin concentrations, we find $\phi \approx 0.9$ and $\ell_m \approx 2a$.
This leads to an estimate for $\Theta \zeta_\perp \tau
\simeq k_B T$ at $310 \, \mbox{K}$ giving $L_p^* \simeq {1
  \over 2}L_p$ and $\ell_c \sim 0.1 \mu \mbox{m}$.

In conclusion, we have studied a simple model of active filaments and
obtained different static and dynamical properties as compared to
passive semiflexible polymers. These differences should be observable
with video microscopy and in linear rheological experiments and 
be relevant for experiments on actin-myosin systems.

We thank A. Ajdari, F. Amblard, E. Furst, D. Humphrey, F. J\"ulicher,
J. K\"as, L. Le Goff and A.C. Maggs for discussions. The financial
support of the Royal Society and the National Science Foundation under
grants PHY-99-07949 (at KITP) is gratefully acknowledged .


\end{multicols}

\begin{references}
  
\bibitem{howard} J. Howard,  {\em Mechanics of Motor Proteins 
and the Cytoskeleton}, (Sinauer, New York, 2000). 
  
\bibitem{nedelec} F. Nedelec et al. {\em Nature} {\bf 389}, 305,
  (1997).
  
\bibitem{kas} D. Humphrey, C. Duggan, D. Saha, D. Smith and J. K\"as,
  {\em Nature} {\bf 416}, 413, (2002).

\bibitem{loic} L. Le Goff, F. Amblard and E. Furst, {\em Phys. Rev. Lett.} 
{\bf 88}, 018101, (2002).

\bibitem{act_mem} J. Prost and R. Bruinsma, {\em Europhys. Lett.}
  {\bf 33}, 321 (1996); J. Prost, J.-B. Manneville and R.  Bruinsma,
  {Eur. Phys. J. B}, {\bf 1}, 465 (1998); R. Granek and S.  Pierrat ,
  {\em Phys. Rev. Lett.} {\bf 83}, 875, (1999).

\bibitem{motile}T. B. Liverpool, A. C. Maggs and A. Ajdari {\em Phys.
    Rev. Lett.} {\bf 86}, 4171, (2001).
  
  
\bibitem{doieds} M. Doi and S.F. Edwards, {\em The Theory of Polymer
    Dynamics}, (Clarendon, Oxford, 1992).
  
  
\bibitem{isamag} H. Isambert and A.C. Maggs, {\em Macromolecules} {\bf
    29}, 1036 (1996).
  
  
\bibitem{Chaikin} Chaikin, P.; Lubensky, T. {\em Principles of
    Condensed Matter Physics}, C.U.P.: Cambridge, U.K., 1995.
  
  
\bibitem{everaers} R. Everaers et al, {\em Phys. Rev. Lett.} {\bf 82},
  3717, (1999).

\bibitem{morse} D.C. Morse, {\em Macromolecules} {\bf 31}, 7030 (1998);
  {\em Macromolecules} {\bf 31}, 7044 (1998); {\em Phys. Rev. E.}
  {\bf 58}, R1237 (1998).
  
\bibitem{freds} F. Gittes and F.C. MacKintosh, {\em Phys. Rev. E.}
  {\bf 58}, R1241 (1998).
  

\bibitem{livmag} T.B. Liverpool and A.C. Maggs, {\em Macromolecules}
  {\bf 34}, 6064-6073 (2001).

\bibitem{morse01} M. Pasquali, V. Shankar, and D. C. Morse  {\em Phys. Rev. E.}
  {\bf 64}, 020802(R) (2001).
  
\bibitem{semiflex} T. Odijk, {\em Macromolecules} {\bf 16}, 1340, (1983);
  M. Doi, {\em J. Polym. Sci.: Polym. Symp.} {\bf 73}, 93,
  (1985); A.N. Semenov, {\em J. Chem. Soc., Faraday Trans.}
  {\bf 2}, 317, (1986).
  
\bibitem{myosin} R.D. Vale and R.A. Milligan, {\em Science} {\bf 288},
  88 (2000); C. Viegel et al, { \em Biophys. J.} {\bf 75}, 1424 (1998). 
  
\bibitem{gardiner} C.W. Gardiner, {\em Handbook of Stochastic Processes}, (Springer, Berlin , 1985).

\end{references}
\end{document}